\documentclass{article}
\usepackage{frascatiphys}
\usepackage{graphicx}
\usepackage{amsfonts,amsmath,amssymb,bm,eucal,latexsym, mathrsfs}
\usepackage{subfig}
\begin{document}
\title{ 
Holographic thermalisation of strongly-coupled systems 
}
\author{
Floriana Giannuzzi        \\
{\em University of Bari, via Orabona 4, Bari, Italy}
}
\maketitle
\baselineskip=10pt
\begin{abstract}
The thermalisation of a strongly-coupled plasma is studied through the AdS/CFT correspondence. The system starts behaving as in viscous hydrodynamics shortly after the end of the perturbation. Local and nonlocal probes are used to characterise the process towards equilibrium.
\end{abstract}
\baselineskip=14pt

\section{Introduction}
One of the challenges of the ALICE experiment is the study of ultrarelativistic heavy ion collisions (HIC) at LHC with the aim of understanding what happens to matter under extreme conditions, similar to those just after the Big Bang, and what are the features of the system.
It has been stated that almost 1 fm/c after the collision the system enters a local equilibrium phase, known as quark-gluon plasma (QGP), and subsequently it expands and cools down until its temperature falls below the QCD transition to the hadronic phase \cite{Heinz:2004pj,Shuryak:2014zxa}.
The QGP contains deconfined partons, and behaves as a nearly perfect fluid in which matter flows collectively, hence it can be described by hydrodynamics at low viscosity. 
However, the description of the system immediately after the collision is much more difficult, since it involves the study of an out-of-equilibrium problem.
Recently, the AdS/CFT correspondence has been applied to such issues \cite{Hubeny:2010ry}, with the aim of understanding how equilibrium is reached and how much time is needed, and of identifying some features of the medium produced in HIC experiments \cite{Chesler:2008hg,Heller:2011ju,Heller:2012km,Wu:2011yd}. 
The  AdS/CFT correspondence, unlike other nonperturbative approaches, can be straightforwardly applied to systems under extreme conditions and far from equilibrium. It has predicted a very small value for the ratio of shear viscosity over entropy density ($\eta/s$), \emph{i.e.} $\eta/s=1/(4 \pi)$, a general result that holds for any strongly-coupled gauge theory having a gravity dual \cite{Policastro:2001yc}.
Experiments at RHIC and LHC have indeed found out that $\eta/s$ should be small in the QGP phase, at odds with predictions of models based on perturbative calculations, thus showing that the AdS/CFT correspondence could capture the strong-coupling features of the system.
Moreover, we  expect that the results obtained through the correspondence for $\CMcal{N} = 4$  super Yang-Mills theory may be relevant to the QGP since it is deconfined and strongly interacting. 

\section{AdS/CFT correspondence in a nutshell}
The AdS/CFT correspondence establishes a duality between a superstring theory in AdS$_5\times S_5$ and a gauge theory, namely $\CMcal{N}=4$ super Yang-Mills theory, living in a four-dimensional (4$d$) Minkowski space \cite{Maldacena:1997re}. 
An interesting aspect of the correspondence is its weak version, according to which the supergravity limit is dual to the large $N$ and strong-coupling limit of the gauge theory, and this suggests that the nonperturbative regime of the gauge theory can be studied by a semiclassical theory. 

The $5d$ AdS (anti-de Sitter) space has a boundary, which is equivalent to a $4d$ Minkowski space (plus a point at infinity), so the dual gauge theory can be built on the boundary of the AdS space. The metric of the AdS$_5$ space in Eddington-Finkelstein coordinates is:
\begin{equation}
\label{eq:EFgeneral}
ds^2=2 dr dt-r^2 dt^2+ r^2 d\vec x^2\,,
\end{equation}
where $r$ is the fifth coordinate, and the boundary is reached in the limit $r\to \infty$.

A specific dictionary has been introduced  to relate the two theories \cite{Witten:1998qj,Gubser:1998bc}. It states that there is a correspondence between a local gauge-invariant operator of the gauge theory and a field in the $5d$ theory, whose mass is related to the conformal dimension $\Delta$ of the operator.
The boundary value of the field is identified with the source of the operator. 
Finally, the generating functional of the correlators of the gauge theory is equal to the partition function of the gravity theory. 

Finite temperature effects can be achieved by introducing a black hole in the $5d$ metric as follows:
\begin{equation}
\label{eq:EFBH}
ds^2=2 dr dt-r^2 (1-r_H^4/r^4) dt^2+ r^2 d\vec x^2\,,
\end{equation}
where $r_H$ represents the position of the horizon of the black hole, which is related to the temperature by
\begin{equation}
\label{eq:rH}
T = \frac{r_H}{\pi}\,. 
\end{equation}
It turns out that at small temperatures the horizon of the black hole is far from the boundary, while at high temperatures the horizon comes close to the boundary. 

\section{Out-of-equilibrium nonabelian conformal plasma}
Let us exploit the AdS/CFT correspondence to study the thermalisation process of an out-of-equilibrium system, as the one produced in HIC.
We require: boost invariance along the collision axis ($x_3$), translation invariance and $O(2)$ rotation invariance in the orthogonal plane $x_\perp=\{x_1,x_2\}$.
It is thus convenient to change coordinates and define the proper time ($\tau$) and the rapidity ($y$) from the time $t$ and $x_3$ by $t=\tau \cosh y$ and $x_3=\tau \sinh y$.
In this approach, a system can be taken out of equilibrium by perturbing the $4d$ metric \cite{Chesler:2008hg}:
\begin{equation}
\label{eq:4dperturbed}
ds^2=-d\tau^2+e^{\gamma(\tau)} dx_\perp^2+\tau^2e^{-2\gamma(\tau)} dy^2\,.
\end{equation}
The metric now contains a factor, $\gamma(\tau)$, which modifies the Minkowski metric for a short time interval until it vanishes or becomes constant. 

Once the $4d$ metric is fixed, Einstein equations have to be solved to fix the corresponding $5d$ metric, which, in general form, can be written as:
\begin{equation}
\label{eq:5dperturbed}
ds^2=2 dr d\tau-A(\tau,r) d\tau^2+ \Sigma(\tau,r)^2 e^B(\tau,r) dx_\perp^2+ \Sigma(\tau,r)^2 e^{-2B(\tau,r)}dy^2\,.
\end{equation}
As a boundary condition, the $5d$ metric must coincide with the $4d$ perturbed metric \eqref{eq:4dperturbed} on the boundary, \emph{i.e.} in the limit $r\to\infty$.
At initial time we require that the metric matches the unperturbed AdS one.

After solving Einstein equations (for the numerical procedure see Ref.~\cite{Bellantuono:2015hxa}), a general result has been obtained: for any considered $\gamma(\tau)$, the $5d$ metric contains a black hole, so, as soon as the perturbation starts, a horizon appears in the $5d$ space. 

The analysis of thermalisation is put forward by means of two kinds of $4d$ observables, \emph{i.e.} local and nonlocal probes. To this aim, an interesting local probe is the energy-momentum tensor, gathering information on energy density and pressure of the system, while nonlocal probes are the equal-time two-point correlation function, and the expectation value of Wilson loops of different shapes, rectangular and circular.
By comparing such observables with their values foreseen in viscous hydrodynamics, it is possible to study thermalisation as the onset of the hydrodynamic regime.

\subsection{Local probes}
The boundary energy-momentum tensor is obtained as the operator dual to the metric tensor
\begin{equation}
\label{eq:EMT}
T^\mu_\nu=\frac{N_c^2}{2 \pi^2}\, \mathrm{diag}(-\epsilon, p_\perp,p_\perp, p_\parallel)\,.
\end{equation}
Its components are the energy density, pressure in the transverse direction with respect to the collision axis, and longitudinal pressure. It can be computed following a recipe based on holographic renormalisation \cite{Bellantuono:2015hxa,deHaro:2000vlm,Kinoshita:2008dq}.
We have also defined an effective temperature using Eq.~\eqref{eq:rH}.

These observables have been computed both with the perturbed metric \eqref{eq:5dperturbed} and with a metric describing a setup governed by viscous hydrodynamics. 
In particular, viscous effects have been included up to second order in an expansion at late time \cite{Heller:2007qt,Baier:2007ix}, obtaining:
\begin{align}
\epsilon_H(\tau)& = \frac{3 \pi^4 \Lambda^4}{4 (\Lambda \tau)^{4/3} }\left[ 1-\frac{2c_1}{ (\Lambda \tau)^{2/3}}+\frac{c_2}{ (\Lambda \tau)^{4/3}} + \CMcal{O}\left( (\Lambda \tau)^{-2} \right )\right] \label{eq:Eidro}\\
p_{\parallel,H} (\tau)& =\frac{ \pi^4 \Lambda^4}{ 4(\Lambda \tau)^{4/3} } \left[ 1-\frac{6c_1}{ (\Lambda \tau)^{2/3}}+\frac{5c_2}{ (\Lambda \tau)^{4/3}} + \CMcal{O}\left( (\Lambda \tau)^{-2} \right )\right] \label{eq:PPAidro}\\
p_{\perp,H} (\tau)& = \frac{ \pi^4 \Lambda^4}{ 4(\Lambda \tau)^{4/3} } \left[ 1-\frac{c_2}{ (\Lambda \tau)^{4/3}} + \CMcal{O}\left( (\Lambda \tau)^{-2} \right ) \right] \label{eq:PPEidro}\\
T_{eff,H}(\tau)& =\frac{\Lambda}{(\Lambda \tau)^{1/3}} \Bigg[ 1-\frac{1}{6 \pi (\Lambda \tau)^{2/3}}+\frac{-1+\log 2}{36 \pi^2 (\Lambda \tau)^{4/3} } + \CMcal{O}\left( (\Lambda \tau)^{-2} \right )\Bigg]
\label{eq:Tidro}
\end{align}
with $c_1=\frac{1}{3 \pi}$, $c_2=\frac{1+2 \log{2}}{18 \pi^2}$; the effective temperature in Eq.~\eqref{eq:Tidro} has been defined from the relation $\epsilon_H=\frac{3}{4}\pi^4 T_{eff,H}^4$.
The parameter $\Lambda$ can be fixed by matching the hydrodynamic temperature with the one computed in the perturbed model at late times.

\subsection{Nonlocal probes}
Nonlocal observables can probe deeper into the bulk spacetime, since they are sensitive to a wide range of energy scales in the boundary field theory, giving a scale dependence of thermalisation.

The equal-time two-point correlation function of an operator with large conformal dimension can be computed from the length $\CMcal{L}$ of the extremal string connecting the points on the boundary \cite{Balasubramanian:2011ur,NLProbes}: 
\begin{equation}
\label{eq:correlator}
\langle \CMcal{O} (t,\bm{x}) \CMcal{O} (t,\bm{x}') \rangle \simeq e^{-\Delta\, \CMcal{L}}\,.
\end{equation}
The expectation value of a spatial Wilson loop can be computed from the Nambu Goto action $\CMcal{S}_{NG}$, which is the area of the surface spanned by the extremal string attached to the contour $\CMcal{C}$ \cite{Balasubramanian:2011ur,NLProbes}:
\begin{equation}
\label{eq:WL}
\langle W_{\CMcal{C}} \rangle \sim e^{-\CMcal{S}_{NG}}\,.
\end{equation}
Details on the definition of these quantities can be found in  \cite{NLProbes}.
Nonlocal probes can have different sizes, corresponding to the distance between the two points on the boundary for the correlation function and to the size of the loop $\CMcal{C}$. 
It is worth emphasising that the strings from which Eqs.~\eqref{eq:correlator}-\eqref{eq:WL} are computed start on the boundary of the $5d$ space and then get into the bulk, to smaller and smaller values of $r$ according to their size: observables with a large size on the boundary can probe deeper into the bulk.
Nonlocal probes have been computed both with the nonequilibrium metric \eqref{eq:5dperturbed} and with a $5d$ metric reproducing, through  holographic renormalisation, the energy-momentum tensor in Eqs.~\eqref{eq:Eidro}-\eqref{eq:PPEidro}.

\section{Results}
\begin{figure}
\begin{center}
\subfloat{
\includegraphics[width = 6.cm]{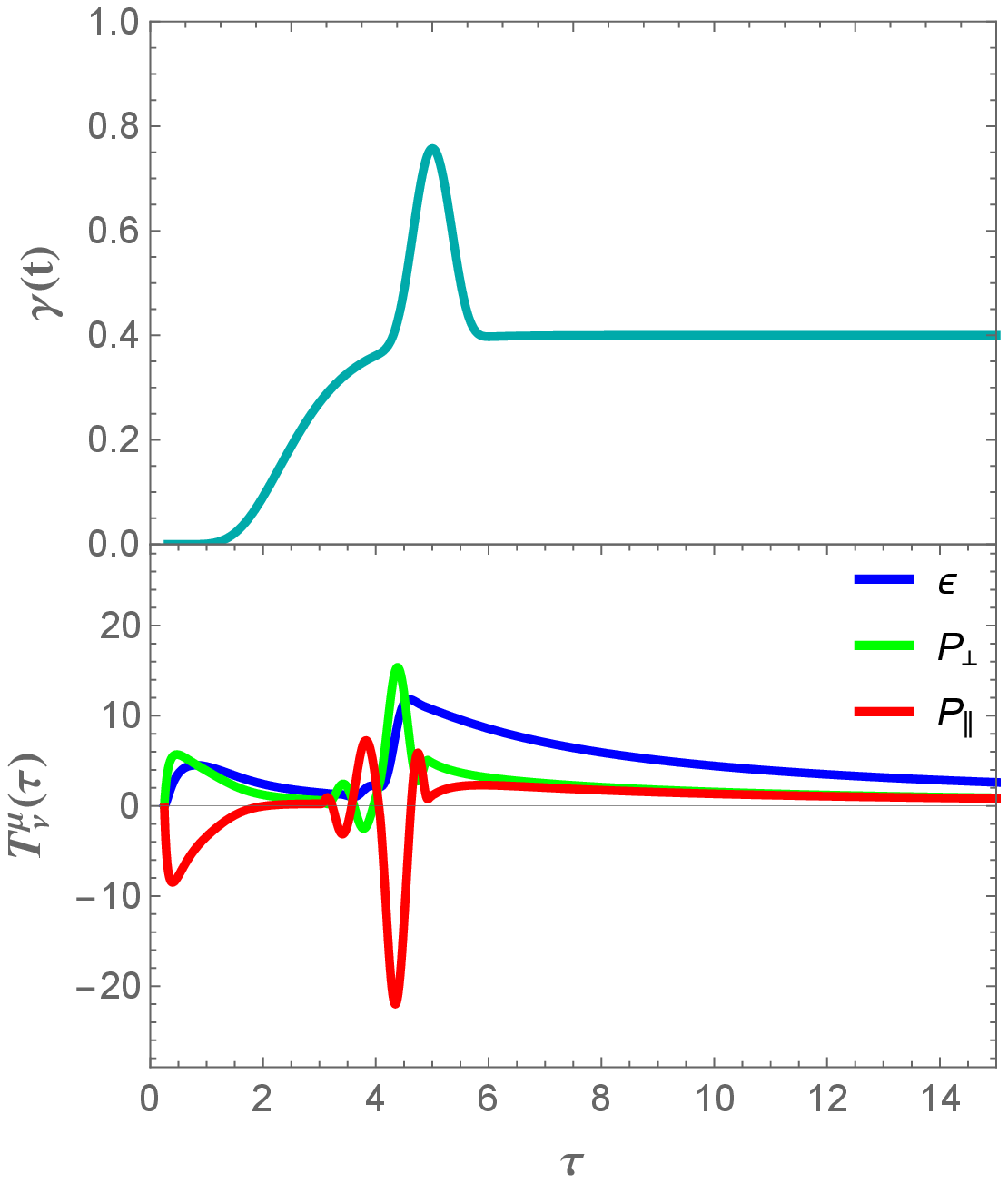}
}
\hspace{1cm}
\subfloat{
\includegraphics[width = 6.cm]{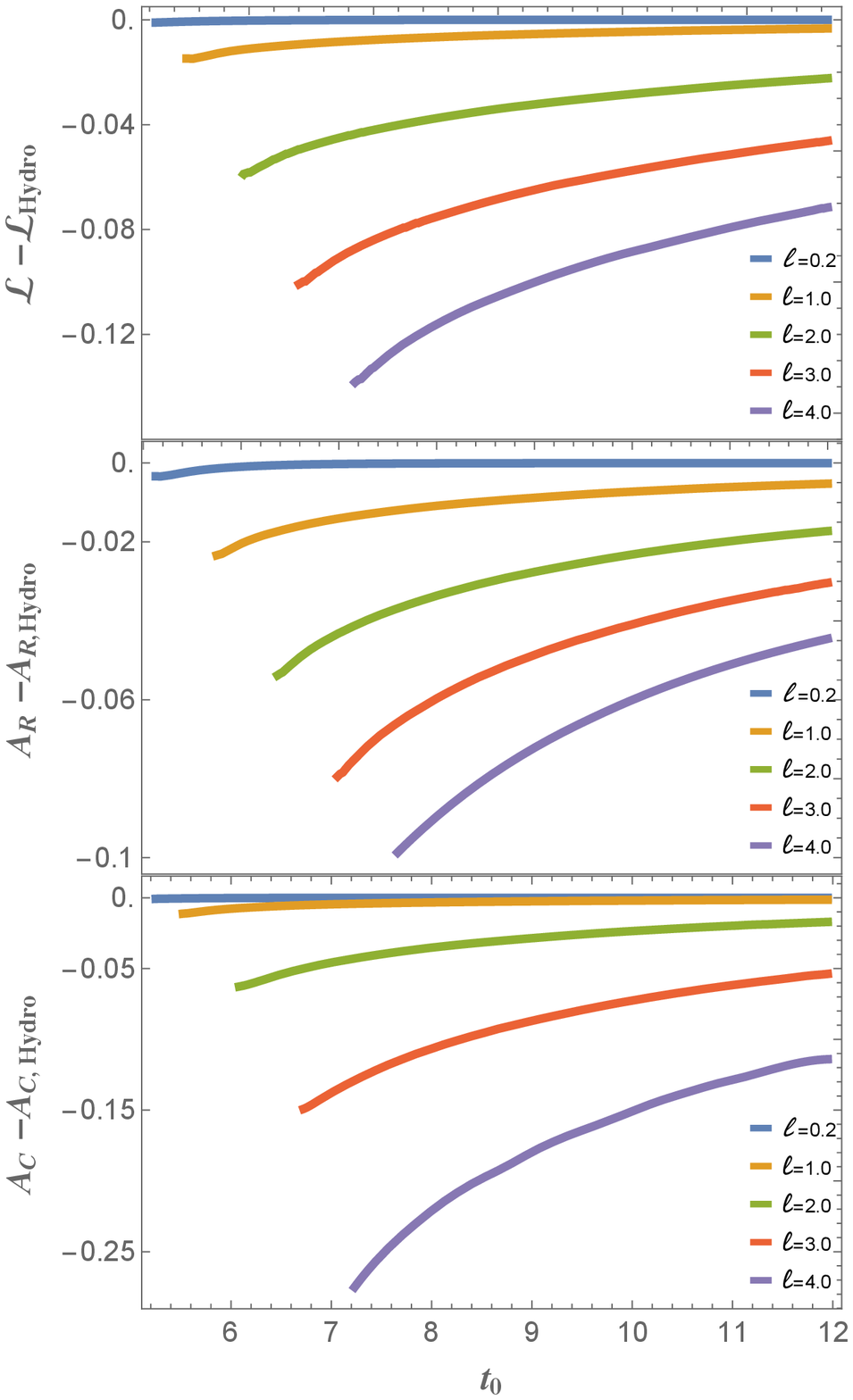}
}
\caption{Left panel: $\gamma(\tau)$ (top), energy-momentum tensor in the out-of-equilibrium model (bottom). Right panel: geodesic length (top), area of the extremal surface for the rectangular Wilson loop (middle), area of the extremal surface for the circular Wilson loop (bottom), computed in the out-of-equilibrium model minus the same quantity in the viscous hydrodynamic model. In the hydrodynamic model the value $\Lambda=1.12$ has been used.}
\label{fig:colonnaEMT}
\end{center}
\end{figure}

Let us adopt the following profile for the function $\gamma(\tau)$ appearing in the $4d$ metric \eqref{eq:4dperturbed} \cite{Bellantuono:2015hxa}:
\begin{equation}
\gamma(\tau) = \frac{2}{5} \tanh^7\left(\frac{\tau-0.25}{1.2}\right)+(1-(\tau-4)^2)^6 e^{-1/(1-(\tau-4)^2)} \Theta(1-(\tau-4)^2) \,.
\label{eq:modB}
\end{equation}
It is characterised, as shown in Fig.~\ref{fig:colonnaEMT}, by one peak plus a slow deformation, becoming at $\tau_f=5$ almost constant (end of the perturbation).
The results obtained for the energy-momentum tensor in the whole time interval are shown in the left panel of Fig.~\ref{fig:colonnaEMT}, while the results for nonlocal probes after the end of the perturbation are in the right panel. 
It emerges that nonlocal probes with larger sizes thermalise later.

\begin{figure}
\begin{center}
\includegraphics[width = 7cm]{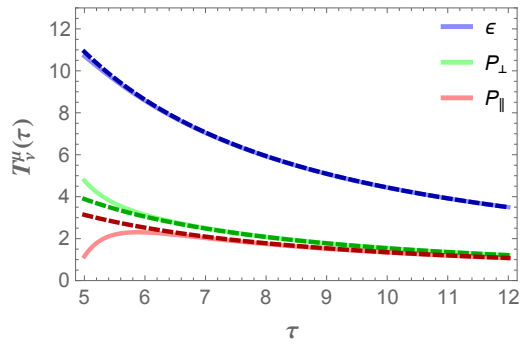}
\caption{Comparison between the energy-momentum tensor in the out-of-equilibrium model (plain curves) and Eqs.~\eqref{eq:Eidro}-\eqref{eq:PPEidro} with $\Lambda=1.12$ (dashed curves).}
\label{fig:EMTzoom}
\end{center}
\end{figure}

It is interesting to look more in detail at what happens to the energy-momentum tensor after the end of the perturbation, as shown in Fig~\ref{fig:EMTzoom}. One can notice that the energy density starts following the hydrodynamic behaviour as soon as the perturbation ends, while longitudinal and transverse pressures are, for a short time interval, quite different from hydrodynamics. 
%

Let us define an equilibration time $\tau^*$ from the relation:
\begin{equation}\label{eq:eqtime}
\Big|\frac{\epsilon(\tau^*)-\epsilon_H(\tau^*)}{\epsilon(\tau^*)}\Big|=0.05\,,
\end{equation}
and an isotropisation time $\tau_p$ from:
\begin{equation}\label{eq:eqtime}
\Big|\frac{p_{\parallel}(\tau_p)/p_\perp(\tau_p)-p_{\parallel,H}(\tau_p)/p_{\perp,H}(\tau_p)}{p_{\parallel}(\tau_p)/p_\perp(\tau_p)}\Big| =0.05\,.
\end{equation}
It turns out that the equilibration time coincides with the end of the perturbation ($\tau^*=5$), while the isotropisation time gets a higher value, $\tau_p=6.74$. 
If we fix an energy scale requiring that the temperature at the end of the perturbation is equal to 500 MeV, we find that complete thermalisation is reached  almost 0.42 fm/c after the end of the perturbation.

The calculation was repeated with  different choices of $\gamma(\tau)$, getting the same qualitative results and a time delay for thermalisation of almost 1 fm/c in all the considered cases \cite{Bellantuono:2015hxa}.

In conclusion, by exploiting the AdS/CFT correspondence we have been able to study an  out-of-equilibrium process, finding  some common features in all the considered models \cite{Bellantuono:2015hxa,NLProbes}, meaning that  the response of the system to perturbation seems to be general.
A first observation is that thermal equilibration and isotropisation are not simultaneous,  the former occurring before the latter.
Full hydrodynamic behaviour is reached after a time of a few fm/c, in agreement with the experimental results.
Another evidence is that local modes equilibrate first, so they need a lower time to thermalise with respect to nonlocal observables with large sizes, confirming that the considered system is strongly coupled.
This approach has been also used to compute other quantities characterising the QGP, like, for example, the energy loss of a heavy quark moving in the plasma \cite{Chesler:2013urd} and the dissociation time of a heavy quark-antiquark pair \cite{Bellantuono:2017msk}.

%

%

\section{Acknowledgements}
I would like to thank L. Bellantuono, P. Colangelo, F. De Fazio, and S. Nicotri for collaboration.

\end{document}